\documentclass[sigconf,natbib=true,anonymous=false]{acmart}
\AtBeginDocument{%
  }
\usepackage{algorithm}
\usepackage{algorithmic}
\usepackage{multirow}

\setcopyright{acmlicensed}
\copyrightyear{2025}
\acmYear{2025}

\begin{document}

\title{DARTS: A Dual-View Attack Framework for Targeted Manipulation in Federated Sequential Recommendation}


\author{qitao qin}
\affiliation{%
  \institution{University of Science and Technology of China}
  \city{hefei}
  \country{China}}
\email{qqt@mail.ustc.edu.cn}

\author{yucong luo}
\affiliation{%
  \institution{University of Science and Technology of China}
  \city{hefei}
  \country{China}
}
\email{prime666@mail.ustc.edu.cn}

\author{zhibo chu}
\affiliation{%
 \institution{University of Science and Technology of China}
 \city{hefei}
 \country{China}}
\email{zb.chu@mail.ustc.edu.cn}

\begin{abstract}
  Federated recommendation systems (FedRec) safeguard user privacy through decentralized training of tailored models, yet they remain susceptible to adversarial threats. Extensive studies have explored targeted attacks in FedRec, driven by commercial and social incentives, but the varying robustness of recommendation algorithms has often been underexplored. Our empirical analysis further indicates that conventional targeted attack strategies yield suboptimal performance in Federated Sequential Recommendation (FSR) scenarios. To address this gap, we investigate targeted attacks in FSR and introduce a novel dual-perspective attack framework, termed DARTS. This approach integrates a sampling-driven explicit attack mechanism with a contrastive learning-guided implicit gradient technique to execute a synergistic attack. Additionally, we propose a specialized defense strategy designed to counter targeted attacks in FSR, enabling a robust evaluation of our attack framework’s mitigation. Comprehensive experiments on benchmark sequential models demonstrate the efficacy of our proposed methodology.
  Our codes are publicly available\footnote{https://anonymous.4open.science/r/DARTS/}.
\end{abstract}
\maketitle

\section{Introduction}
The rapid advancement of recommendation systems has significantly enhanced their performance, yet it has also amplified concerns regarding user privacy. Centralized training in conventional recommendation systems poses risks of substantial financial losses and critical privacy breaches in the event of data compromise.
\begin{figure}
    \includegraphics[scale=0.26]{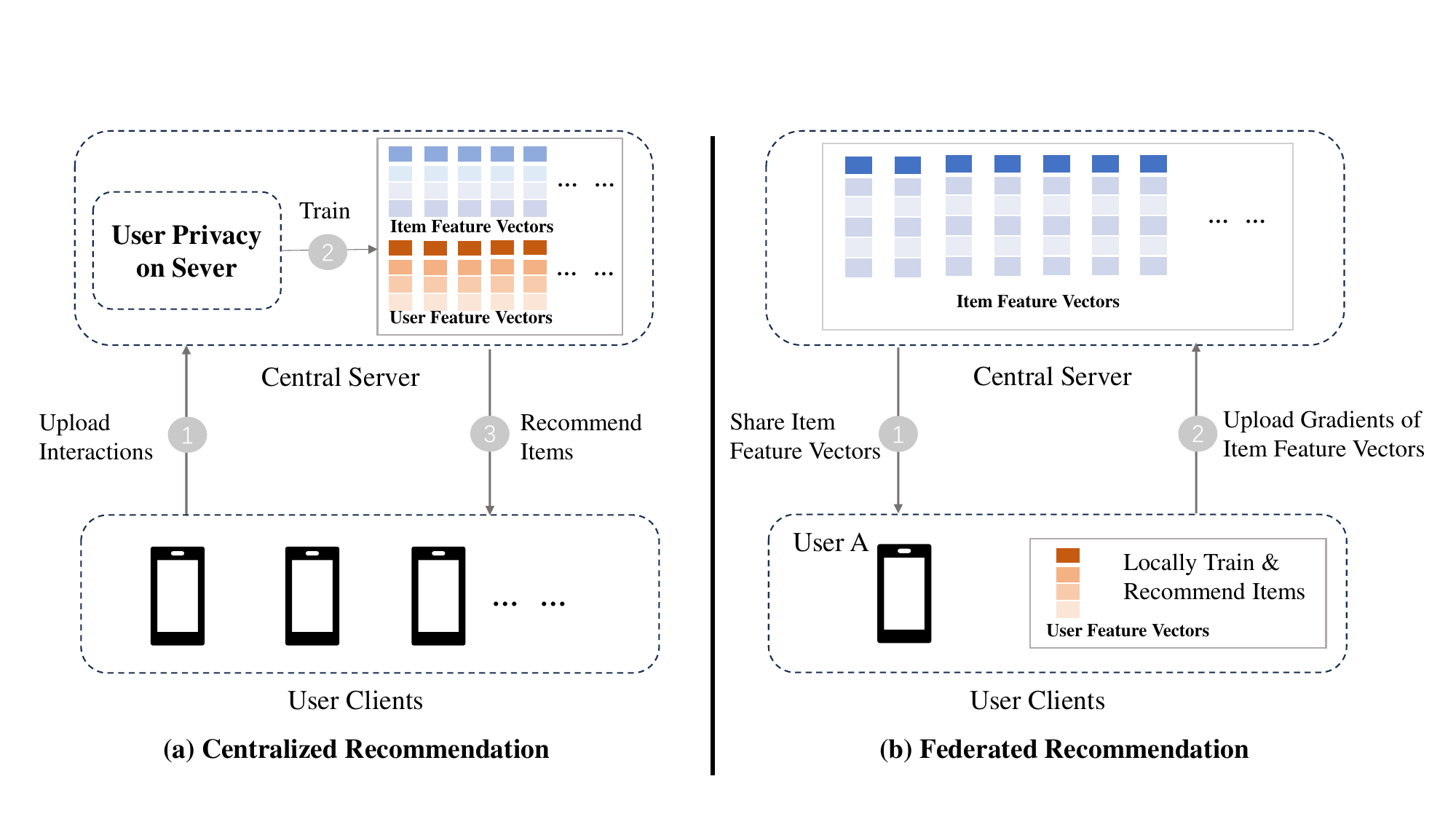}
    \caption{Centralized v.s. Federated Recommendation}
    \label{fig:enter-label-1}
    \Description{}
\end{figure}
To mitigate these challenges, federated recommendation systems have gained traction as an effective privacy-preserving approach~\cite{ammad2019federated,lin2020fedrec,liang2021fedrec++}. As illustrated in Figure 1, these systems safeguard user data by decentralizing the training process, enabling users to train models locally on their devices and subsequently upload model gradients to a central server. Despite the privacy enhancements offered by federated learning (FL), vulnerabilities persist. The decentralized nature of the training process~\cite{mcmahan2017communication} allows malicious actors to tamper with local training data or manipulate uploaded gradients to pursue harmful objectives~\cite{bhagoji2019analyzing, bagdasaryan2020backdoor}.


Thereby, over the past years, some attack methods aimed at federated recommendation systems have emerged. Based
on the goal of the attacker, poisoning attacks can be classified into targeted and untargeted attacks.
The targeted attacks aims to boost targeted item exposure, in several cases,  it can even influence public opinion on social media and change the public's views or attitudes on specific heat events, resulting substantial commercial and social impact.
However, in studies focusing on the robustness of federated recommendation systems especially in targeted item attack~\cite{rong2022fedrecattack,zhang2022pipattack,rong2022poisoning}, experiments on sequential recommendation models are almost nonexistent. Instead, nearly all experimental efforts are concentrated on Matrix Factorization (MF)~\cite{li2016data} and 
Neural Collaborative Filtering (NCF)~\cite{he2017neural} models. Moreover, our empirical findings in Table 1 indicate
that existing targeted attack methods achieve only limited effectiveness in  federated sequential recommendation (FSR) tasks. With an increasing number of studies~\cite{li2022federated, shi2024privacy, zhang2024feddcsr} in recent years linking sequential recommendations to FL in pursuit of data privacy, it is necessary to explore the security of  FSR.


The targeted poisoning attack against  FSR faces several critical challenges. (1) 
The attack must be effective even with a small fraction of malicious clients and limited access to data. Considering that recommender systems usually have millions of users, it is impractical for the attacker to control a large number of clients. Moreover, as clients in Federated Recommendation Systems (FSR) do not share their local training data, the attacker can only access a small set of data stored on the malicious clients.
(2) 
Existing federated targeted attack methods often rely on certain global information, such as a proportion (e.g., 10\%)~\cite{rong2022fedrecattack} of the training data or item popularity rankings~\cite{zhang2022pipattack}. However, such data is typically inaccessible to attackers. Therefore, we impose a restriction on the use of global information, which presents an even greater challenge for our attack.
(3) 
Many recommender systems are trained on implicit user feedback, which inherently contains significant noise, making them naturally resistant to a certain extent against malicious perturbations~\cite{wang2021learning}.

To address these challenges, in this paper, we first propose a novel dual-view targeted model poisoning attack method named DARTS. 
Our main idea is to fully leverage the interaction sequences in sequential recommendations from multiple perspectives. 
First, from an explicit perspective, we identify and replace the item that has the greatest influence on the prediction of the target item, uploading malicious gradients to boost the target item’s score. Then, from an implicit perspective, we use contrastive learning to strengthen the similarity between the embeddings of the target item and the interacted items. Even with a small number of malicious users, no access to global data, and the model's inherent robustness due to noise in its training on implicit feedback, our approach proves to be both adaptable and effective.

As the significant threat posed by this dual-view targeted attack, even when only a small number of users are compromised, we seek to explore defense mechanisms to mitigate this attack effect. We drew inspiration from the concept of the geometric median and proposed a hybrid robust weight aggregation algorithm to mitigate this formidable target item attack. 

The main contributions of our work are listed as follows:
\begin{itemize}
    \item To our knowledge, this study is among the first to explore the robustness of sequential recommendation models in the context of federated target item attacks. Our findings suggest that existing attack methods may have limited effectiveness when applied to FSR.
    \item We propose DARTS, a novel dual-view targeted poisoning attack method, which reveals the security risk of FSR.
    \item To evaluate the mitigation effects of the DARTS attack, we propose the mixed-RFA defense strategy.
    \item  Extensive experiments on two public datasets and models validate the effectiveness of our attack and defense. 
\end{itemize}

\section{Related Work}
In this section, we briefly introduce the related work from two categories: attack and defense on federated recommendations, sequential recommendation systems.
\subsection{Attack \& Defense on Federated Recommendation}
Poisoning attacks against recommender systems and their defense have been widely studied in the past decades~\cite{lam2004shilling,aktukmak2019quick}. However, these researches mainly focus on the centralized training of recommendation models. They require the attacker or the server to have strong knowledge of the full training data to perform effective attacks or defenses, such as all users profiles~\cite{burke2006classification} and the entire rating matrix~\cite{li2016data}. These methods are infeasible under the federated learning (FL) setting since the server cannot access the data of clients. The defense of recommendation mainly include adversarial training~\cite{tang2019adversarial}, knowledge distillation~\cite{papernot2016distillation}, and attack detection~\cite{lee2012shilling}. However, virtually all defense methods for recommendation systems have not been applied in federated recommendation systems. 

In the general FL domain, recently, several untargeted attacks have been proposed to degrade the overall performance of recommendation systems~\cite{yu2023untargeted, wu2022fedattack, yi2023ua}. Simultaneously, several targeted poisoning attack methods have been proposed and can be directly applied to boost certain target items in FedRec scenario. For example, zhang~\cite{zhang2022pipattack} proposes the method to improve the target item score and popularity by designing new loss function. Rong~\cite{rong2022fedrecattack} needs a certain proportion of public interactions to increase the target item score. Rong~\cite{rong2022poisoning} proposes an model poisoning function using Binary Cross-Entropy (BCE) loss and approximating user's embedding vectors. 
\begin{figure*}
    \includegraphics[scale=0.41]{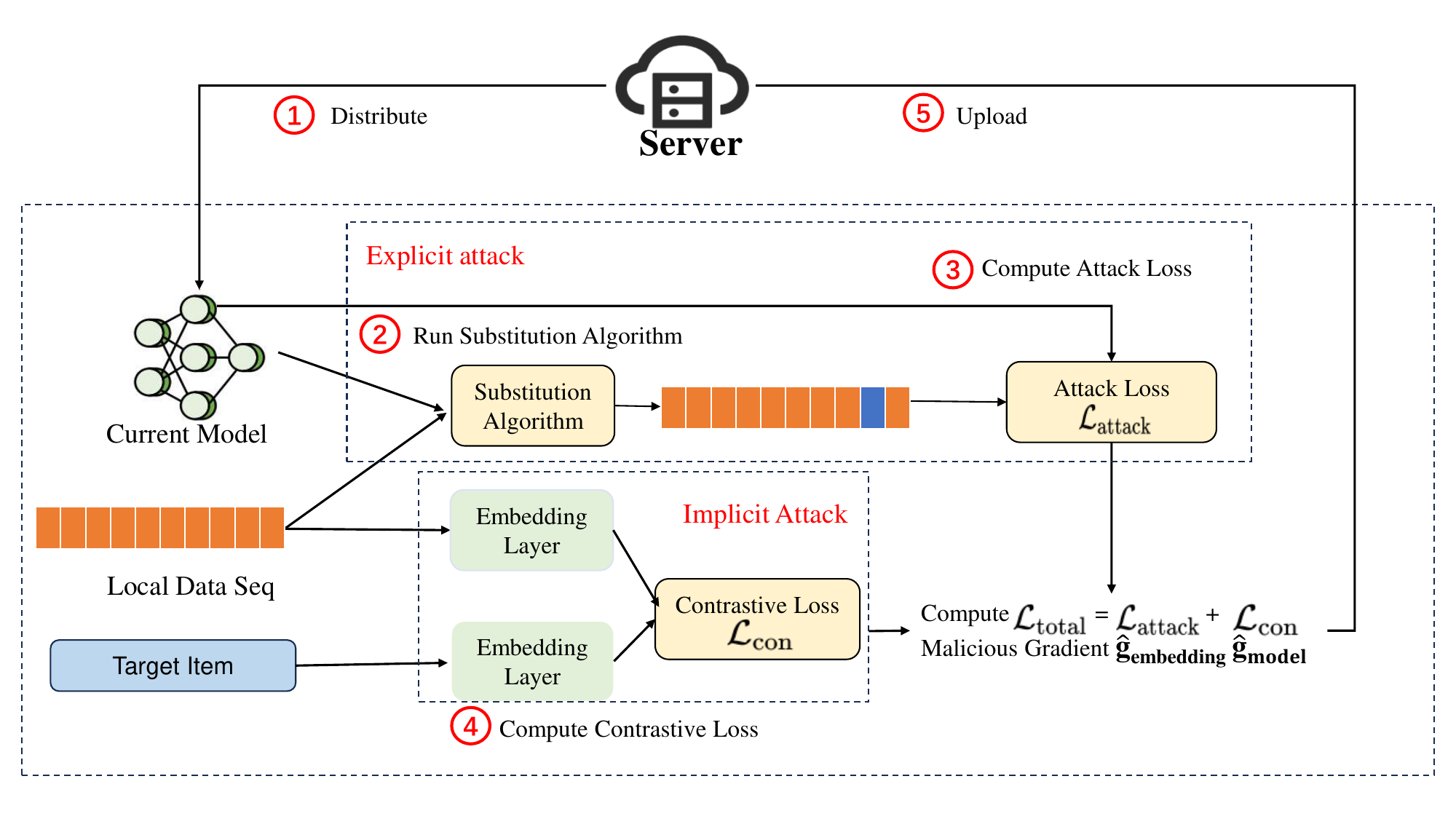}
    \caption{The Procedure of Our DARTS Attack}
    \label{fig:enter-label}
    \Description{}
\end{figure*}
\subsection{Sequential Recommendation}
Sequential recommendation systems~\cite{liu2022one} focus on modeling the sequential behavior of users as they interact with items over time. It primarily consists of two components: user-item interactions~\cite{cheng2021learning} and sequence model~\cite{cheng2022towards}. It exploits the interaction sequences between users and items to reveal the users' next behavior~\cite{fang2020deep, quadrana2018sequence, wang2021survey}. The attention mechanism can boost the sequential recommendation performances~\cite{li2017neural, liu2018stamp}. For instance, the SASRec model employs a two-layer Transformer~\cite{vaswani2017attention} decoder to model user sequence, and BERT4Rec utilizes a bidirectional method to model the users' sequential behaviors. But All of these methods adopt centralized training methods, which may cause the risk of data privacy leakage.

Despite the extensive research on attacks in the past, the robustness of recommendation models has been overlooked. In this paper, we focus on target item attacks in FSR.

\section{Preliminaries}

In this section, we briefly introduce the framework of federated sequential recommendation (FSR) and threat model.
\subsection{Federated Sequential Recommendation}
Let $I$ and $U$ denotes the sets of M items and N users (clients) in a recommender system, respectively. These clients try to train a global model collaboratively without sharing their private data. We denote the sequential recommender with function $f$. Given an input sequence $x$, $f$ predicts a probability distribution over the item scope $I$. The function $f$ comprises an embedding function $f_e$ and a sequential model $f_m$, with $f(x) = f_m(f_e(x))$. For a data pair $(x, y)$, ideally, $f$ predicts $y$ with the highest probability (i.e., $y = \arg\max f(x)$). The learning of a sequential recommender $f$ is to maximize the probability of output item $y$ upon input $x$. In other words, we minimize the expectation of loss $L$ with respect to $f$ over the data distribution $X$:
\begin{equation}
\min_f \mathbb{E}_{(x,y) \sim X} L(f(x), y) \tag{1}
\end{equation}
where $L$ represents the training loss function (i.e., ranking loss or cross entropy loss).
In FSR, every user's item interaction is separated. Our sequential model is distributedly trained under the framework of federated recommendation. More specially, in federated learning (FL) scenarios, there is a central server and a large amount of individual user clients. As each user corresponds to one of the user clients, we use user to represent its client for convenience. In sequential recommendation, we assume that it all consists three parts: an item interaction sequence $ x $,  embedding function $f_e$ and sequential model $f_m$. In each training round, the server first distributes the current global parameters [$f_e$;$f_m$]  to $n$ randomly selected clients. Then each selected client computes the update gradient g =  [$g_{embedding}$; $g_{model}$] with their local sequence $x$. Following previous work~\cite{rong2022poisoning}, we use Binary Cross-Entropy (BCE)  loss to train the local model. Specially, the gradient g is generated by optimizing the following loss function:
\begin{equation}
\mathcal{L} = -\frac{1}{|D|} \left( \sum_{(x_t, y_t) \in D^+} \log \hat{y}_t + \sum_{(x_t, y_t) \in D^-} \log (1 - \hat{y}_t) \right) \tag{2}
\end{equation}
$D^+$ represent all interacted items collection, $D^-$ represents all non-interacted items collection, $\hat{y}_t$ represent the probability of model predicted sample $x_t$.  
Next, the client uploads [$g_{embedding}$; $g_{model}$] to the server. It is important to note that the user interaction sequence can not be uploaded due to its privacy sensitivity. Finally, the server aggregates all the received gradients with certain aggregation rules and updates the global model. Such training round proceeds iteratively until convergence. 
\subsection{Threat Model}
\paragraph{Attack Goal.} The attacker aims to increase the target item's exposure $\epsilon$ of the FSR system on arbitrarily inputs.
\paragraph{Attack Capability and knowledge.} The attacker controls a set of malicious clients $\mathcal{U}_{mal}$ which accounts for $m$\% of $\mathcal{U}$. As there are usually millions of users in a recommender system, we assume that $m$ should be small (e.g., $m = 0.1$). Following previous works~\cite{wu2022fedattack, zhang2022pipattack, yu2023untargeted}, we assume that the attacker has access to the training code, local model, and user data on the devices of malicious clients while cannot access the data or gradients of other benign clients. The attacker can arbitrarily modify the gradients uploaded by the malicious clients. We also assume the attacker does not know the aggregation rule used by the server.

\section{Methodology}

In this section, we first provide a detailed description of the newly proposed DARTS framework. Subsequently, we also introduce a newly proposed defense mechanism.
\subsection{DARTS Framework Overview}
Following the previous works~\cite{yu2023untargeted,rong2022poisoning}, we backdoor the target recommender model similarly by uploading a mix of clean gradients and poisoned gradients. Let $\tilde{\mathcal{L}}$ denote the loss function which can indicate the goal of our attack. In the t-th round of training, each selected benign user $\mu$ uploads gradients of $\mathcal{L}_u$(i.e. clean gradients). To achieve the goal of our attack, we manipulate the selected malicious users to upload gradients of $ \tilde{\mathcal{L}}$(i.e. poisoned gradients). The actual loss function of the recommender model under our attack can be represented:
\begin{equation}
\mathcal{L}_{server} = \sum_{u \in \mathcal{U}} \mathcal{L}_u + \alpha \tilde{\mathcal{L}} \tag{3}
\end{equation}
where $\alpha$ is a positive coefficient that trades off between the model validity and the attack effectiveness.
To address the problems, we adopt steps  to design a proper loss function $\tilde{\mathcal{L}}$ for our attacks.
We utilized the algorithm steps in Figure 2 and Algorithm 1 to carry out the dual-view DARTS attack.
When selected for model training, the malicious client receives  the latest global model  from the server, which contains embedding and model (step 1). 
\subsubsection{Explicit Strategy.}To better utilize the sequence, we run Substitution Algorithm, and choose one most vulnerable item to replace (step 2). We provide a detailed description of the Substitution Algorithm demonstrated in Algorithm 2.
\begin{itemize}
    \item \textbf{Forward and Compute Gradients}. First, we initialize $x'$ unchanged to $x$, followed by computing the embedded sequence $\tilde{x}'$ using the embedding function $f_e$. Then $\tilde{x}'$ is fed through the recommender function $f_m$ to compute the cross entropy loss $L_{ce}$ w.r.t target item t. Then, backward propagation is performed to retrieve the gradients $\Delta_{\tilde{x}'}$(i.e. $\Delta_{\tilde{x}'}$ = $\Delta_{\tilde{x}'} L_{ce}(f_m(\tilde{x}',t))$. 
    \item \textbf{Select Vulnerable Items to Attack}. In input sequence, items are of different vulnerability. We select only one vulnerable item and perform substitution on such items to achieve the best attack performance. We calculate importance scores using $\Delta_{\tilde{x}'}$ from the previous step. We choose the first one item from importance ranking \textbf{r}. Based on the fast gradient sign method, the perturbed embedding $\tilde{x}'$ can be computed.
    \item  \textbf{Project Embedding to Item}. We project the perturbed embedding $\tilde{x}'$ back to the item, we compute cosine similarity between $\tilde{x}'$ and candidate items from $\mathcal{I}$, items with higher similarity values are favored as adversial substitutes. Here, we impose another constraint to enforce item similarity, where adversarial items are required to have a minimum cosine similarity ($\tau$) with the original items.  We will select from the top $T$ items with the highest cosine similarity, and replace the one item that predicts the highest score for the target.
\end{itemize}
After that, following previous work~\cite{rong2022poisoning},  instead of  maximizing $\sum_{i \in \tilde{I}} \epsilon_i$, we can change the goal of our attacks to maximizing $\frac{1}{|U|} \sum_{i \in \tilde{I}} \sum_{u \in U} \hat{Y}_{ui}$. $\tilde{I}$ represents target item set. So we follow previous works, use BCE Loss to maxmize target item's score and get $L_{attack}$ (step 3):
\begin{equation}
    L_{\text{attack}} = \text{BCE}(f(s')[t], y) \tag{4}
\end{equation}
It is noted that if the target item and test item in history interaction, it does not appear in next item prediction. 

\begin{algorithm}
    \caption{DARTS Attack}
    \begin{algorithmic}[1]
        \STATE \textbf{Input:} Sequence ${x}$, Target ${t}$, Recommendation Model  $f_{s}(i.e.,f_e,f_m)$ 
        \STATE \textbf{Output:} malicious gradients $\hat{g}_{embedding}, \hat{g}_{model}$
        \\ // Step 1:  distribute model
        \STATE Initialize \hspace{-0.2em} client model $f_i(f_e,f_m) \gets $$f_{s}(f_e,f_m)$
        \\ // Step 2: run Substitution Algorithm
        \STATE Compute Substituted Sequence in Algorithm 2 \\ \hspace{0.5cm}$x' \gets substitution(f_i(f_e,f_m),x,t)$
        \\ // Step 3: compute attack loss
        \STATE Compute $L_{\text{attack}} \gets \text{BCE}(f(\text{s'})[\text{t}], y) $
        \\ // Step 4: compute contrastive  loss
        \STATE Randomly sample n non-interacted item i. Compute negative vector $\mathbf{n_{i}} \gets f_e(i)$
        \STATE Compute positive vector and anchor $\mathbf{p} \gets mean(f_e(x))$ \hspace{0.1cm}
        $\mathbf{a} \gets f_e(t)$ 
        \STATE Compute $L_{con}$ by equation (7)
        \\ // Step 5: compute total loss and malicious gradient
        \STATE Compute $L_{total}$,   $\hat{g}_{embedding}, \hat{g}_{model}$ by equation (8)
        \\ // Step 6: Upload gradients
        \STATE Upload $\hat{g}_{embedding}, \hat{g}_{model}$  to server
    \end{algorithmic}
\end{algorithm}
\subsubsection{Implicit Strategy.}Inspired by recent works on contrastive learning, we introduced an embedding contrastive loss to increase the score of the target item (step 4). This approach aims to enable the target item to learn useful features from interacted items. We use target item embedding as anchor $\mathbf{a}$, the mean of interacted item embedding as positive vector $\mathbf{p}$, randomly sample $n$ non-interacted items as negative vector $\mathbf{n}$ (e.g. $\mathbf{n_{1},n_{2}...}$) and use the cross-entropy loss function for loss calculation. Finally, we get $L_{con}$. The specific contrastive learning calculation formula is shown as follows:
\begin{itemize}
    \item \textbf{Cosine Similarity Calculation:}
    \begin{itemize}
        \item Positive similarity: anchor and positive sample:
        \begin{equation}
        s_{\text{pos}} = \frac{\mathbf{a} \cdot \mathbf{p}}{\|\mathbf{a}\| \|\mathbf{p}\|}\tag{5}
        \end{equation}
        where $\mathbf{a}$ is anchor vector and $\mathbf{p}$ is positive vector.
        \item Negative similarities between the anchor and each negative sample:
        \begin{equation}
            s_{\text{neg}_i} = \frac{\mathbf{a} \cdot \mathbf{n}_i}{\|\mathbf{a}\| \|\mathbf{n}_i\|} \tag{6}
        \end{equation}
        where $\mathbf{n}_i$ represents each negative sample vector.
    \end{itemize}


    \item \textbf{Softmax Cross-Entropy Loss:}
    The target label for the anchor is always the positive sample (index 0), so the cross-entropy loss function is:
    \begin{equation}
    L_{con} = -\log \left( \frac{e^{s_{\text{pos}}}}{e^{s_{\text{pos}}} + \sum_{i=1}^{n} e^{s_{\text{neg}_i}}} \right) \tag{7}
    \end{equation}
\end{itemize}

\begin{algorithm}  
    \caption{Substitution Algorithm}
    \begin{algorithmic}[1]
        \STATE \textbf{Input:} Sequence ${x}$, Target ${t}$, Recommendation Model $f(i.e.,f_e,f_m)$, Search Time $T$ (default $T = 9$)
        \STATE \textbf{Output:} Sequence ${x}'$ with only one item replaced
        \STATE Initialize polluted sequence $ x'\gets x $;
        \STATE Compute sequence embeddings ${\tilde{x}' \gets f_e(x')}$;
        \STATE Compute gradients w.r.t. $\tilde{x}'$:\\ $\Delta_{\tilde{x}'}\gets \Delta_{\tilde{x'} } L_{ce}(f_m(\tilde{x}'),t)$;
        \STATE \hspace{0cm}select one most vulnerable items in $x'$: \\ \hspace{0.5cm}$i \gets argmax(||\Delta_{\tilde{x}'}||)$
        \STATE Compute cosine similarity $s$: 
        \[
        s \gets \text{CosSim}\left(\tilde{x}_i' - \text{sign}(\nabla_{\tilde{x}_i'}), f_e(c)\right) \quad \forall c \in \mathcal{I}
        \]
        \STATE \text{Impose similarity constraint } $s_c:$ \\ 
        \hspace{0.5cm}$s_c \gets \mathbf{1}\left(\text{CosSim}(\mathbf{x}_i', f_e(\mathbf{c})) \geq \tau\right) \quad \forall \mathbf{c} \in \mathcal{I}$
        \STATE Update cosine similarity $s$: $s \gets s \odot s_c$
        \STATE Compute similarity ranking indice:\\ 
        \hspace{0.5cm}$s_{indice} \gets argsort(s,descending)$
        \STATE Set target scores $target_{scores} \gets -np.inf$\\ 
        Copy $x'$ to $x''$ $x''\gets x'$
        \FOR{$j$  in $T$}
        \STATE Replace $x_i'' \hspace{0.2cm} in \hspace{0.2cm} x'': {x}_i'' \gets s_{indice}[j]$  
        \IF{$f(x'')[t] > target_{scores}$}
        \STATE {$ {x}_i' \gets s_{indice}[j]$} 
        \STATE$target_{scores} \gets f(x'')[t] $
        \ENDIF
        \ENDFOR
        
    \end{algorithmic}
\end{algorithm}
The final loss function formula is as follows (step 5):
\begin{equation}
    L_{\text{total}} = L_{\text{attack}} + L_{\text{con}}\tag{8}
\end{equation}
Finally, compute the gradients of the embedding  $\hat{g}_{embedding}$ and model $\hat{g}_{model}$  based on the loss function, and upload them to the server(step 5 \& 6).

\subsection{Defense with Mixed-RFA}

In federated learning systems, particularly those applied to recommendation systems, robust aggregation strategies are crucial for performance and security. The geometric median (GM) of vectors \( w_1, \ldots, w_m \) in \( \mathbb{R}^d \) with weights \( \alpha_1, \ldots, \alpha_m > 0 \) provides a robust statistic, as it is defined as the minimizer \( v \) of the function:
\begin{equation}
g(v) = \sum_{i=1}^m \alpha_i \|v - w_i\| \tag{9}
\end{equation}
The minimization of this function leads to a central point that minimizes the weighted Euclidean distance to all other points in
the dataset, thereby enhancing the system’s resistance to outliers and adversarial attacks that might skew the aggregated results.

Despite its robustness, the geometric median often results in a significant degradation in the performance of recommendation systems. This degradation can occur because the GM may overly penalize the system’s ability to leverage informative but non-central local models, which are crucial for capturing diverse user preferences and behaviors effectively. To address these limitations, we propose a novel hybrid federated aggregation algorithm that com-
bines the robustness of the geometric median with the performance efficiency of traditional averaging methods. The proposed method
can be formalized as follows
To address these limitations, we propose a novel hybrid federated aggregation algorithm that combines the robustness of the geometric median with the performance efficiency of traditional averaging methods. The proposed method can be formalized as follows:
\begin{equation}
v = \lambda \left( \sum_{i=1}^m \alpha_i w_i \right) + (1-\lambda) \cdot \text{GM}(w_1,..., w_m; \alpha_1,..., \alpha_m) \tag{10}
\end{equation}
Here, \( \lambda \) is a tuning parameter that balances the influence of the simple weighted average and the geometric median in the final aggregation. By adjusting \( \lambda \), the system can be tailored to prioritize either robustness or performance according to the specific needs and threat models of the federated environment.

\section{Experiments}
\begin{table}[ht]\small
\captionsetup{position=bottom, skip=10pt}
\centering
\begin{tabular}{lcccc}
\toprule 

\textbf{Dataset} & \textbf{\#Users} & \textbf{\#Items} & \textbf{\#Interactions} & \textbf{\#Sparsity} \\
\midrule 
ML-1M & 6,040 & 3,706 & 1,000,209 & 95.53\% \\
Steam & 3,753 & 5,134 & 114,713 & 99.40\% \\
\bottomrule 

\end{tabular}
\caption{ Sizes of Datasets} 
\label{table:dataset-sizes}
\vspace{-0.3cm}
\end{table}
In this section, we conduct several experiments to answer the following research questions(RQs):
\begin{itemize}
    \item \textbf{RQ1:} How is the effectiveness of our attack compared to that of existing attacks?
    \item \textbf{RQ2:} Does the architecture of different models significantly impact their susceptibility to attacks in federeated sequential recommendation? 
    \item \textbf{RQ3:} How does the ratio of malicious clients affect the performance of our attack methods?
    \item \textbf{RQ4:} How does our attack method DARTS perform under defensive mechanisms mixed-RFA?
    \item \textbf{RQ5:} Are the implicit and explicit strategy in our attack method truly effective? 
\end{itemize}
\subsection{Datasets and Experiment Setting}
\paragraph{Dataset.} 
For our experimental analysis, we selected two authentic real-world datasets representing markedly distinct domains: movie recommendation and game recommendation. These datasets include \textbf{MovieLens-1M} (abbreviated as ML-1M)~\cite{harper2015movielens} and \textbf{Steam-200K} (referred to as Steam)~\cite{cheuque2019recommender}, with their respective scales and characteristics detailed in Table 2. To ensure consistency with established methodologies in the field, we implemented a leave-one-out evaluation strategy, where the most recent item interacted with by each user is reserved as the test set, while the penultimate item is utilized for the training phase. This approach allows us to assess model performance under realistic conditions, reflecting the dynamic nature of user interactions across diverse recommendation contexts. By incorporating these varied datasets, we aim to capture a broad spectrum of user behaviors and preferences, thereby enhancing the generalizability of our findings.
\paragraph{Experiment Setting.} 
In our experimental setup, we opted for two well-established recommendation models, SASRec and BERT4Rec, which are widely recognized for their effectiveness in sequential recommendation tasks. The federated learning (FL) framework we implemented is based on the FedAvg algorithm~\cite{mcmahan2017communication}, a cornerstone in distributed learning systems. Within the FedRec system, we designate each user as an independent client to reflect real-world federated scenarios. To simulate adversarial conditions, we randomly sampled a subset of users from the complete user pool U, specifically selecting 0.1\%, 0.2\%, 0.3\%, 0.4\%, 0.5\%, and 1\% of users, and designated them as malicious clients to introduce perturbations into the system. In alignment with the methodology outlined in~\cite{rong2022fedrecattack}, we evaluated the recommendation models’ performance using two key metrics: the \textit{Hit Ratio (HR)} and the \textit{Normalized Discounted Cumulative Gain (NDCG)}, both computed over the top 10 ranked items for the test item, providing a robust assessment of ranking quality. To gauge the success of targeted item attacks, we employed the average \textit{Exposure Ratio at K (ER@K)} for the target items as our primary evaluation metric. To thoroughly explore the attack dynamics across varying scales, we configured \textit{K} at multiple levels, specifically 5, 10, 20, and 30, allowing us to capture the nuanced impact of attack intensity on system performance. Additionally, for the defense mechanism \textit{mixed-RFA}, which aims to mitigate adversarial effects, we tuned the hyperparameter $\lambda$ to 0.3, balancing robustness and computational efficiency.
\begin{table*}[h]
\captionsetup{position=bottom, skip=10pt}
\centering
\renewcommand\arraystretch{1.4}
\tabcolsep=0.3cm
\footnotesize
\scalebox{1.1}{
{\begin{tabular}{cccccccccc}
\hline
\textbf{Model} & \textbf{Dataset} & \textbf{Attack} & \textbf{Mal. Client} & \textbf{HR@10}  & \textbf{NDCG@10} & \textbf{ER@5} & \textbf{ER@10} & \textbf{ER@20} & \textbf{ER@30} \\ \hline
\multirow{12}{*}{SASRec} & \multirow{6}{*}{ML-1M} & None & 0 & 0.1061 & 0.0505 & 0.0000 & 0.0000 & 0.0000 & 0.0021 \\ 
&  & RA & 0.1\% & \textbf{0.1061}  & \textbf{0.0508} & 0.0000 & 0.0000 & 0.0000 & 0.0035 \\
&  & EB & 0.1\% & 0.1038  & 0.0499 & 0.0023 & 0.0161 & 0.1068 & 0.2916 \\
&  & A-ra & 0.1\% & 0.1050  & 0.0499 & 0.0028 & 0.1062 & 0.1111 & 0.3031 \\\cline{3-10}
&  & \textbf{\multirow{2}{*}{DARTS}} & 0.1\% &  0.0929   & 0.0372  & \textbf{0.8890} & \textbf{0.9981} & \textbf{1.0000} & \textbf{1.0000} \\
&  &  & 0.05\% &  0.0969   & 0.0421 & 0.6701 & 0.8951 &  1.0000 & 1.0000 \\\cline{2-10}
& \multirow{6}{*}{Steam} & None & 0 & 0.3141  & 0.1546 & 0.0000 & 0.0000 & 0.0000 & 0.0000 \\
&  & RA & 0.1\% & \textbf{0.3307}  & \textbf{0.1732} & 0.0000 & 0.0000 & 0.0000 & 0.0036 \\
&  & EB & 0.1\% & 0.3109  & 0.1591 & 0.0000 & 0.0003 & 0.0144 & 0.1148 \\
&  & A-ra & 0.1\% & 0.3131  & 0.1590 & 0.0000 & 0.0003 & 0.0160 & 0.1156 \\\cline{3-10}
&  & \textbf{\multirow{2}{*}{DARTS}} & 0.1\% & 0.2944 & 0.1411 & \textbf{0.7169} & \textbf{0.9997} & \textbf{1.0000} & \textbf{1.0000} \\
&  &  & 0.05\% & 0.3131 & 0.1623 & 0.0011 & 0.0117 & 0.2950 & 0.8311 \\\cline{1-10}
\multirow{12}{*}{BERT4Rec} & \multirow{6}{*}{ML-1M} & None & 0 & 0.1174  & 0.0564 & 0.0000 & 0.0000 & 0.0000 & 0.0000 \\
&  & RA & 1\% & 0.1184  & 0.0567 & 0.0000 & 0.0000 & 0.0000 & 0.0000 \\
&  & EB & 1\% & 0.1172  & 0.0571 & 0.0018 & 0.0086 & 0.0526 & 0.1879 \\
&  & A-ra & 1\% & \textbf{0.1189}  & \textbf{0.0577} & 0.0012 & 0.0084 & 0.0553 & 0.1930 \\\cline{3-10}
&  & \textbf{\multirow{2}{*}{DARTS}} & 1\% & 0.1058  & 0.0422  & \textbf{0.9531} & \textbf{0.9872} & \textbf{1.0000} & \textbf{1.0000} \\
& &  & 0.2\% & 0.1132 & 0.0554 & 0.1152 & 0.4518 & 0.9767 & 1.0000\\\cline{2-10}
& \multirow{6}{*}{Steam} & None & 0 & 0.3293  & 0.1697 & 0.0000 & 0.0000 & 0.0000 & 0.0000 \\
&  & RA & 1\% & \textbf{0.3339}  & \textbf{0.1737} & 0.0000 & 0.0000 & 0.0000 & 0.0000 \\
&  & EB & 1\% & 0.3304  & 0.1700 & 0.0000 & 0.0000 & 0.0037 & 0.1423 \\
&  & A-ra & 1\% & 0.3312 & 0.1705 & 0.0000 & 0.0000 & 0.0035 & 0.1335 \\\cline{3-10}
&  & \textbf{\multirow{2}{*}{DARTS}} & 1\% & 0.3128   & 0.1417  & \textbf{0.9324}  & \textbf{0.9892}  & \textbf{1.0000}  & \textbf{1.0000}\\
&  &  & 0.2\% & 0.3304 & 0.1699  & 0.0059 & 0.1492  & 0.9379  & 1.0000
\\ \hline
\end{tabular}}
}
\caption{Performance Metrics of Different Federated Learning Models Under Various Attack Vectors}

\end{table*}
\paragraph{Baseline Attacks.} Following the common setting for federated recommendation, we assume that attacker does not have any prior knowledge at all. Bandwagon attack~\cite{kapoor2017review}, Pipattack~\cite{zhang2022pipattack} and FedRecAttack rely on side information of items' popularity or public interactions, cluster attack~\cite{yu2023untargeted} and fedAttack~\cite{wu2022fedattack} focus on untargeted items attack, hence they are not applicable under such circumstances. We choose the attacks which are still practical under such circumstances as our baseline attacks:
\begin{itemize}
    \item Random Attack (RA)~\cite{kapoor2017review}. It injects fake users as malicious users, and manipulates them to interact with both target items and randomly selected items.
    \item Explicit Boosting (EB)~\cite{zhang2022pipattack}. EB is one component of PipAttack which does not rely on attacker's knowledge. It explicitly boosts the predicted scores of target items for malicious users.
    \item A-ra~\cite{rong2022poisoning}. A-ra a variant of the original A-ra which does not approximate user's embedding vectors with normal distribution, because in sequential recommendation models, interaction sequences are used instead of user embedding.
\end{itemize}
\begin{table}[h] \footnotesize
\captionsetup{position=bottom, skip=10pt}
\centering
\renewcommand\arraystretch{2}

\setlength{\tabcolsep}{1.5mm}{

\begin{tabular}{cccccc}
\hline
\textbf{Dataset} & \textbf{Model} & \textbf{ER@5} & \textbf{ER@10} & \textbf{ER@20} & \textbf{ER@30} \\ \hline
\multirow{2}{*}{ML-1M}& SASRec & 0.9890 & 0.9981 & 1.0000 & 1.0000 \\ 
& BERT4Rec & 0.0007 & 0.0046 & 0.0396 & 0.1262 \\ \cline{1-6}
\multirow{2}{*}{Steam}& SASRec & 0.9169 & 0.9997 & 1.0000 & 1.0000 \\ 
& BERT4Rec & 0.0000 & 0.0003 & 0.0059 & 0.2190 \\ \cline{1-6}

\end{tabular}
}
\caption{Different Models under 0.1\% Malicious Client}
\end{table}
\subsection{Attack Performance Evaluation (RQ1)}
We conducted an in-depth evaluation to assess the efficacy of baseline attack strategies in comparison with our proposed approach. The outcomes of these experiments are systematically documented in Table 1, revealing several critical insights. Firstly, the naive data poisoning strategy, referred to as RA, proves ineffective due to the extremely low proportion of malicious users within the system, resulting in negligible impact on the target model. Secondly, the baseline model poisoning techniques, namely EB and A-ra, exhibit only marginal enhancements in their attack efficacy. This limited improvement can likely be attributed to their design, which, while tailored for model poisoning, struggles to achieve significant disruption in sequential recommendation systems when the presence of malicious users is minimal, thus failing to effectively undermine the model’s integrity. Thirdly, our innovative dual-view DARTS framework demonstrates a remarkable superiority over the baseline strategies. By strategically minimizing the embedding distance between the target item and the items previously interacted with by the user—while requiring only a single interaction modification—our approach achieves outstanding attack success rates. Furthermore, it is particularly noteworthy that DARTS maintains its high effectiveness even under conditions where user-related information is drastically reduced, as evidenced by the results in the final row of Table 1. This consistent performance highlights the robustness and operational efficiency of DARTS, setting it apart from competing methods and underscoring its potential as a powerful tool in adversarial settings involving federated sequential recommendation systems.
\subsection{Performance Under Attack (RQ2)} 
\begin{figure}[h]
    \centering
    \hspace{-0.5cm}
    \includegraphics[width=1\linewidth]{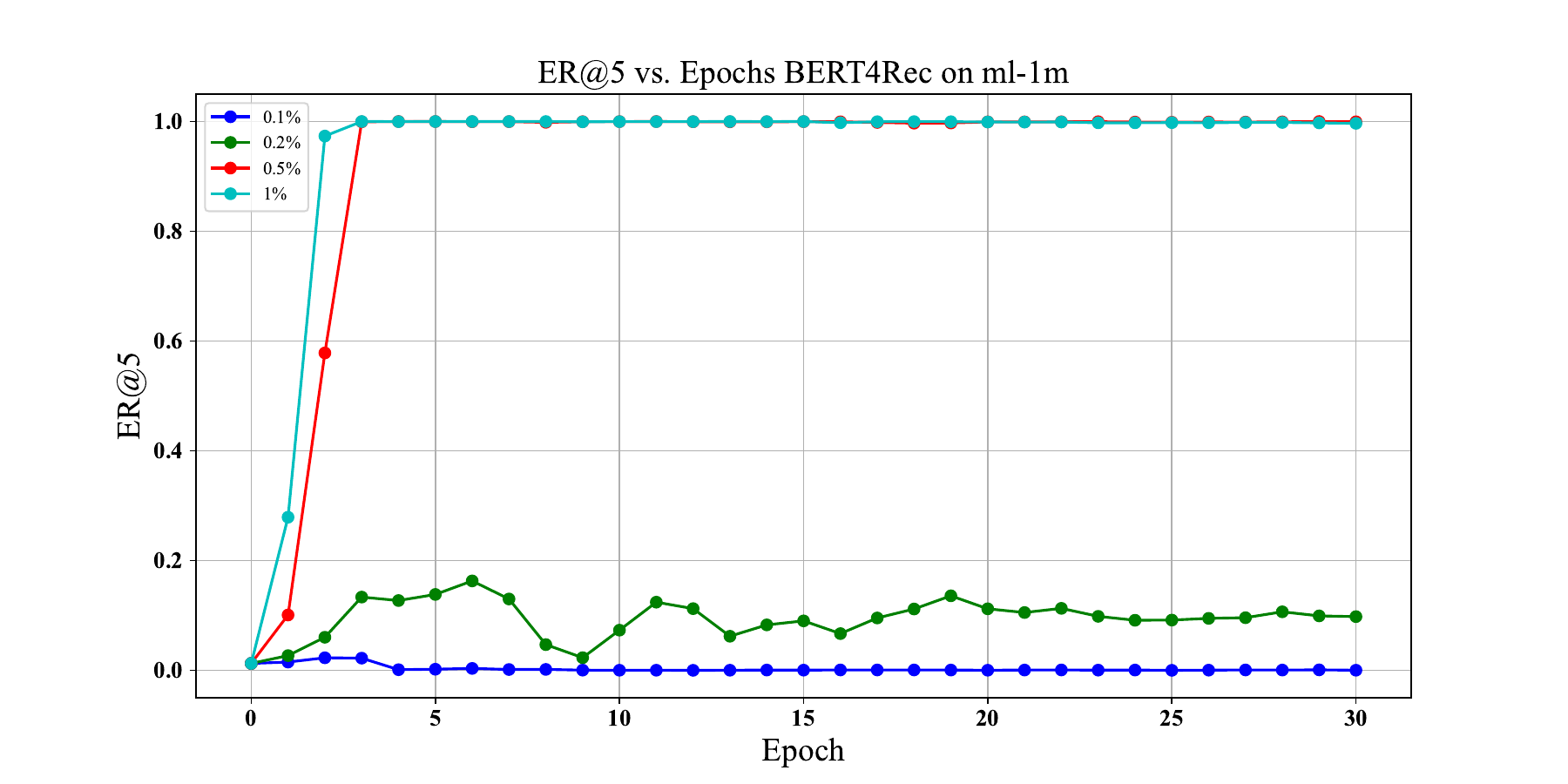}
    \label{fig:bert-ml} 
    \hspace{-0.5cm}
    \includegraphics[width=1\linewidth]{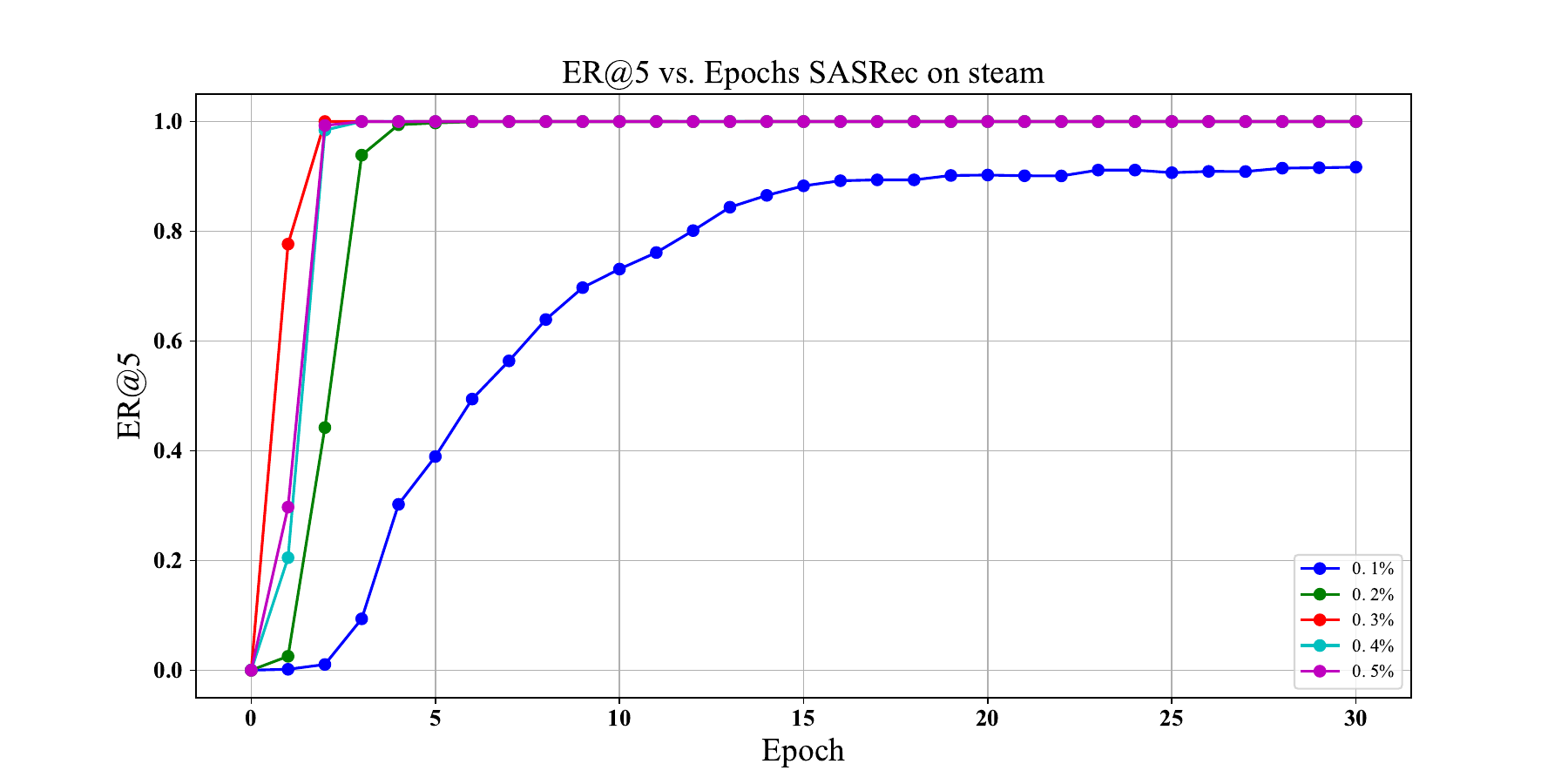}
    
    \caption{Changes in ER@5 Under DARTS}
    \label{fig:sasrec-steam}
    \Description{}
\end{figure}
\begin{table}[]\footnotesize
\captionsetup{position=bottom, skip=10pt}
\centering
\renewcommand\arraystretch{2}
\tabcolsep=0.3cm
\setlength{\tabcolsep}{0.6mm}{
\begin{tabular}{ccccccc}
\hline
\textbf{Dataset} & \textbf{Mal. Client} & \textbf{Defense} & \textbf{ER@5} & \textbf{ER@10} & \textbf{ER@20} & \textbf{ER@30} \\ \hline
\multirow{9}{*}{ML-1M} & \multirow{3}{*}{0.05\%} & none & 0.9701  & 0.9851 & 1.0000  & 1.0000 \\ 
&  & adv\_train & 0.9698  & 0.9838 &  1.0000  & 1.0000   \\
&  & mixed-RFA & \textbf{0.0000}  & \textbf{0.0000}   & \textbf{0.0000}  &   \textbf{0.0014}  \\\cline{2-7}
& \multirow{3}{*}{0.1\%} & none & 0.9890 & 0.9981 & 1.0000 & 1.0000 \\
&  & adv\_train &  0.9834  & 0.9950  & 1.0000  & 1.0000  \\
& & mixed-RFA & \textbf{0.3935} & \textbf{0.8457}  & \textbf{0.9997}  & 1.0000  \\ \cline{2-7}
& \multirow{3}{*}{0.2\%} & none & 1.0000   & 1.0000  & 1.0000  & 1.0000  \\
&  & adv\_train &  1.0000  & 1.0000  & 1.0000  & 1.0000  \\
&  & mixed-RFA & 1.0000   &1.0000  & 1.0000  & 1.0000  \\ \cline{1-7}
\multirow{9}{*}{Steam}  & \multirow{3}{*}{0.05\%}& none & 0.0011  & 0.0117  & 0.2950  & 0.8311 \\
&  & adv\_train & 0.0000  &  0.0080 &  0.2241 &  0.7445 \\
&  & mixed-RFA & \textbf{0.0000}  &  \textbf{0.0000} & \textbf{0.0000}  & \textbf{0.0005}  \\\cline{2-7}
& \multirow{3}{*}{0.1\%} & none & 0.9169  & 0.9997 & 1.0000  & 1.0000  \\
&  & adv\_train &  0.9217   & 0.9992  & 1.0000   & 1.0000  \\
& & mixed-RFA & \textbf{0.0578}   & \textbf{0.4391} & \textbf{0.9739}  & \textbf{0.9995}  \\ \cline{2-7}
& \multirow{3}{*}{0.2\%} & none & 1.0000   &1.0000  & 1.0000   & 1.0000   \\
&  & adv\_train &  1.0000   & 1.0000  & 1.0000  & 1.0000  \\
& & mixed-RFA & \textbf{0.9526}  & 1.0000  & 1.0000 & 1.0000  \\ \hline
\end{tabular}
}
\caption{The Effectiveness of Attacks Under Defenses} 
\end{table}
\begin{table*}[h]
\captionsetup{position=bottom, skip=10pt}
\centering
\renewcommand\arraystretch{1.8}
\tabcolsep=0.3cm
\footnotesize
\begin{tabular}{cccccccccc}
\hline
\textbf{Model} &\textbf{Mal.client} & \textbf{Dataset} & \textbf{Attack} & \textbf{ER@5} & \textbf{ER@10} & \textbf{ER@20} & \textbf{ER@30} \\ \hline
\multirow{6}{*}{SASRec} & \multirow{6}{*}{0.05\%} & \multirow{3}{*}{ML-1M} & C-FSR & 0.0005  & 0.0017 & 0.0132  & 0.0257 \\ 
&  &  & S-FSR & 0.5661  & 0.7396 & 0.9653  & 1.0000  \\
&  &  & DARTS & \textbf{0.9701}  & \textbf{0.9851}  & \textbf{1.0000}  & \textbf{1.0000}    \\\cline{3-8}
&  & \multirow{3}{*}{Steam} & C-FSR & 0.0000 & 0.0000 & 0.0006 & 0.0015 \\
&  &  & S-FSR &  0.0000  & 0.0013  & 0.0184 & 0.1010 \\
&  & & DARTS & \textbf{0.0011} & \textbf{0.0117}  & \textbf{0.2950}  & \textbf{0.8311}  \\ \cline{1-8}
\multirow{6}{*}{BERT4Rec} & \multirow{6}{*}{0.2\%} & \multirow{3}{*}{ML-1M}& C-FSR & 0.0000  & 0.0002  & 0.0011  & 0.0158 \\
&  &  & S-FSR &   0.0978   &  0.4081  & 0.9644   & 1.0000   \\
&  &  & DARTS &  \textbf{0.1152}  & \textbf{0.4518}   &  \textbf{0.9767}  &  \textbf{1.0000}  \\\cline{3-8}
&  & \multirow{3}{*}{Steam} & C-FSR & 0.0000   & 0.0008  & 0.0012   & 0.0153   \\
&  &  & S-FSR &  0.0037    & 0.1276   & 0.7233    & 0.9997   \\
&  & & DARTS & \textbf{0.0059 }   & \textbf{0.2492 } & \textbf{0.9379 }  & \textbf{1.0000 }   \\ \hline
\end{tabular}
\caption{Ablation Study for Module in Attack}
\end{table*}
Although both BERT4Rec and SASRec operate as sequential recommendation system models, BERT4Rec exhibits a notable level of resilience that surpasses SASRec under challenging conditions. The detailed experimental outcomes, as outlined in Table 3, reveal distinct patterns of performance when the system is subjected to a minimal presence of malicious users, set at 0.1\%. Even when exposed to our sophisticated attack technique, DARTS, the exposure rate (ER) of SASRec experiences a marked decline, whereas BERT4Rec maintains its performance with virtually no discernible reduction in ER. Further analysis under identical attack conditions indicates that SASRec becomes vulnerable with just 0.1\% malicious users, leading to a substantial rise in the ER for the targeted item. In contrast, BERT4Rec demands a higher threshold, requiring at least 0.5\% or potentially up to 1\% of malicious users to exhibit any noticeable effect on its functionality. This disparity in vulnerability is largely explained by the inherent bidirectional architecture of BERT4Rec, which equips it with a superior capacity to mitigate inaccuracies. By drawing on contextual information from both the preceding and following sequences, BERT4Rec can effectively address gaps or distortions in the input data, thereby enhancing its stability and reliability. This adaptive strength not only underscores its robustness against adversarial manipulations but also suggests its potential applicability in environments where data integrity is frequently compromised, such as federated learning settings with sequential recommendation tasks.

\subsection{More Malicious Users Increase Attack Effectiveness (RQ3)} 

To thoroughly examine the resilience of our targeted attack strategy across a diverse range of conditions, we performed an extensive analysis by applying varying intensities of the DARTS attack, with specific thresholds set at 1\%, 0.5\%, 0.4\%, 0.3\%, 0.2\%, and 0.1\%. This comprehensive evaluation was executed across two distinct datasets, namely ml-1m and steam, to ensure broad applicability. As illustrated in Figure 3, a pronounced pattern becomes evident: as the proportion of malicious users within the network rises, the attack achieves its maximum efficacy at an accelerated pace and maintains this heightened influence over an extended period, with little evidence of waning strength. Additionally, as detailed in Table 1, our attack technique stands out by delivering enhanced performance while relying on a remarkably reduced volume of user data—evidenced by a mere 0.05\% data requirement, in contrast to the 0.1\% demanded by competing approaches. This efficiency not only highlights the resourcefulness of our method but also reinforces its superiority in practical deployment. The observed benefits are likely tied to the innovative design of DARTS, which optimizes resource allocation and enhances adaptability, making it particularly suited for dynamic environments where data availability fluctuates, such as federated recommendation systems.

\subsection{Defense Performance Evaluation (RQ4)}
In this section, we systematically analyze the protective capabilities of our mixed-RFA defense mechanism when applied to the SASrec model. To establish a benchmark, we employ a well-established sequential recommendation defense technique, known as adv\_train~\cite{yue2022defending}, which is designed to counteract gradient-based training attacks. While the adv\_train approach has previously delivered impressive results in defending against adversarial threats within traditional sequential recommendation frameworks, its performance notably deteriorates when deployed in federated learning environments, where data privacy and decentralization introduce additional complexities. To address this limitation, we introduce a novel strategy that leverages a mixed robust gradient ensemble to counteract targeted attacks on specific items. The experimental results, as presented in Table 4, indicate that our defense mechanism successfully neutralizes targeted item attacks when the fraction of malicious users remains minimal, such as at levels of 0.05\% and 0.1\%. 
Nevertheless, as the proportion of malicious users escalates to 0.2\%, the effectiveness of our defense approach begins to wane, eventually rendering it insufficient to thwart adversarial actions. This 
observation underscores the critical importance of safeguarding user privacy in such systems. Should a substantial amount of user data be exposed, even the most carefully designed defense mechanisms may struggle 
to prevent sophisticated marketing-oriented attacks launched by adversaries, highlighting the need for robust privacy-preserving techniques in federated learning contexts.

\subsection{Ablation Study (RQ5)}
We have crafted multiple adaptations of the initial DARTS approach by strategically removing selected elements, enabling a thorough investigation into their individual influence on the overall success of the attack mechanism. The first adapted version, designated as C-FSR, deliberately excludes the explicit strategy outlined in step 3, whereas the second version, labeled S-FSR, skips the implicit strategy detailed in step 2. These deliberate alterations provide a structured opportunity to dissect and assess the specific contributions of each element within the DARTS architecture. The findings from our comprehensive ablation analysis, meticulously recorded in Table 5, reveal that the original DARTS configuration consistently achieves superior results compared to its modified counterparts, C-FSR and S-FSR, across a range of malicious user percentages. This consistent dominance suggests that the integration of both explicit and implicit strategies is essential for optimizing the accuracy and overall influence of the attack. Furthermore, these insights shed light on the intricate dynamics of item ranking manipulation, offering practical guidance for refining attack strategies in diverse recommendation system environments, particularly those involving federated learning frameworks.
\section{Conclusion}

In this research, we delved into the challenges associated with executing targeted attacks against Federated Sequential Recommendation (FSR) systems, with a primary focus on the notable variations in resilience among different recommendation models. To tackle this complex issue, we introduced a novel dual-view targeted attack framework, dubbed DARTS, which skillfully integrates both explicit and implicit tactics to execute synchronized and potent attacks. Through a series of comprehensive experiments, we established that DARTS effectively undermines the system's defensive capabilities. These findings not only validate the efficacy of our approach but also reveal critical vulnerabilities that can be exploited within federated environments. We aspire that this study will ignite broader curiosity and stimulate deeper investigations into model-specific attack strategies tailored for federated recommendation platforms, thereby fostering the creation of advanced techniques to bolster security measures and protect against emerging threats in such distributed systems.
\newpage

\appendix

\section{Experimental Setting}

In our experimental design, we configured the hidden dimensions of both the SASRec and BERT4Rec models to a uniform value of 64, ensuring consistency across the architectures. To prevent overfitting and enhance model generalization, we established a dropout ratio of 0.1 as a regularization technique. The batch\_size across all experimental runs was fixed at 256 to maintain uniformity in processing large datasets efficiently. Additionally, we standardized the maximum sequence length to 200 for both models across the two datasets, facilitating a fair comparison of their performance. Our federated learning (FL) framework relies on the FedAvg algorithm~\cite{mcmahan2017communication}, a widely adopted method for distributed training. Within this federated setup, each user actively contributes by uploading gradient updates at the conclusion of every training round, prompting us to limit the training duration to a maximum of 30 rounds while consistently applying a weight decay strategy to regulate model complexity. For the contrastive learning component, we defined the number of negative samples P as 100 to enrich the learning process with diverse examples. During the evaluation phase of our federated learning configuration, we randomly selected 1,000 negative samples to compute pertinent performance metrics for both the test and target item, ensuring a thorough analysis. It is worth emphasizing that, to avoid bias in predictions, if either the target item or the test item is present within the interaction sequence, they are excluded from the subsequent next-item prediction task.
\section{Experimental Environment}

We conduct experiments on a Linux server with Ubuntu 22.04.4 LTS. All experiments were run on an NVIDIA GeForce RTX 4090 GPU with CUDA 12.1. The CPU is Intel(R) Xeon(R) Gold 6426Y CPU @2.50GHZ. We use Python 3.8.0 and Pytorch 2.3.1.
\begin{table}[h]\footnotesize
\captionsetup{position=bottom, skip=10pt}
\centering
\renewcommand\arraystretch{1.6}
\tabcolsep=0.4cm
\setlength{\tabcolsep}{0.8mm}{
\begin{tabular}{ccccccc}
\hline
\textbf{Dataset} & \textbf{Mal. Client} & \textbf{Attack} & \textbf{ER@5} & \textbf{ER@10} & \textbf{ER@20} & \textbf{ER@30} \\ \hline
\multirow{12}{*}{ML-1M} & \multirow{3}{*}{0.05\%} & EB & 0.0000  & 0.0000  & 0.0000  & 0.0000  \\ 
&  & A-ra & 0.0000  & 0.0000  &  0.0000  & 0.0000  \\
&  & DARTS & 0.0000  & 0.0000   & 0.0000  &   0.0000  \\\cline{2-7}
& \multirow{3}{*}{0.1\%} & EB & 0.0000 & 0.0000 & 0.0000 & 0.0000 \\
&  & A-ra &  0.0000  & 0.0000 & 0.0000  & 0.0000  \\
& & DARTS & \textbf{0.3935} & \textbf{0.8457}  & \textbf{0.9997}  & 1.0000  \\ \cline{2-7}
& \multirow{3}{*}{0.15\%} & EB & 0.0000 & 0.0000 & 0.0000 & 0.0000\\
&  & A-ra &  0.0000  & 0.0000  & 0.0000   & 0.0000  \\
& & DARTS & \textbf{1.0000}  & \textbf{1.0000}  &  \textbf{1.0000}  & \textbf{1.0000}   \\ \cline{2-7}
& \multirow{3}{*}{0.2\%} & EB & 0.0000   & 0.0000 &  0.0000 &  0.0005 \\
&  & A-ra &  0.0000 &0.0000  & 0.0000  & 0.0010 \\
&  & DARTS & \textbf{1.0000}   &\textbf{1.0000} & \textbf{1.0000}  &\textbf{1.0000} \\ \cline{1-7}
\multirow{12}{*}{Steam}  & \multirow{3}{*}{0.05\%}& EB & 0.0000  & 0.0000  & 0.0000  & 0.0000 \\
&  & A-ra & 0.0000  &  0.0000 &  0.0000 &  0.0000 \\
&  & DARTS & 0.0000  &  0.0000 & 0.0000  & \textbf{0.0005}  \\\cline{2-7}
& \multirow{3}{*}{0.1\%} & EB & 0.0000  & 0.0000 & 0.0000  & 0.0000  \\
&  & A-ra &  0.0000   & 0.0000  & 0.0000  & 0.0000  \\
& & DARTS & \textbf{0.0578}   & \textbf{0.4391} & \textbf{0.9739}  & \textbf{0.9995}  \\ \cline{2-7}
& \multirow{3}{*}{0.15\%} & EB &  0.0000 & 0.0000 & 0.0005  & 0.0048 \\
&  & A-ra & 0.0000   & 0.0000  & 0.0005   & 0.0048 \\
& & DARTS & \textbf{0.7146}   & \textbf{0.9877} & \textbf{1.0000}  & \textbf{1.0000}  \\ \cline{2-7}
& \multirow{3}{*}{0.2\%} & EB & 0.0000 &0.0005 &  0.0144  & 0.0991 \\
&  & A-ra &   0.0000  & 0.0005  & 0.0149 &  0.1058\\
& & DARTS & \textbf{0.9526}  & 1.0000  & 1.0000 & 1.0000  \\ \hline
\end{tabular}
}
\caption{Effectiveness of Attacks Under mixed-RFA } 
\end{table}
\section{Supplementary Experiments}
To provide stronger evidence supporting the resilience and efficacy of our proposed attack framework, DARTS, in the context of federated sequential recommendation (FSR), we conducted a detailed examination and systematically documented the ultimate exposure rates resulting from a range of attack techniques functioning under the protective framework termed mixed-RFA. This in-depth evaluation centers on comparing the capabilities of our innovative DARTS approach with those of two widely recognized gradient-based attack techniques, namely A-ra~\cite{rong2022poisoning} and EB~\cite{zhang2022pipattack}. By leveraging this comparative analysis, we aim to highlight the distinct advantages of DARTS, particularly its ability to maintain high performance levels even when confronted with the defensive measures embedded in the mixed-RFA system. This study not only underscores the adaptability of our method across varying adversarial conditions but also sets a foundation for future explorations into optimizing attack strategies within distributed recommendation environments.
Detailed in Table 6, our findings illustrate that under the mixed-RFA, the efficacy of the DARTS method can be significantly curtailed when the proportion of malicious clients remains below the threshold of 0.1\%. This indicates an effective defense response at lower threat levels. However, as the malicious client ratio surpasses this threshold, notably beyond 0.1\%, the DARTS strategy markedly enhances the exposure of the target item. This heightened exposure substantially escalates the risk posed to the security of FSR, showing the critical vulnerabilities in these environments.

In stark contrast, traditional attack methods, including A-ra and EB, exhibit only limited effectiveness even when the malicious client ratio escalates to 0.2\%. This discrepancy underscores the superior adaptability and potency of the DARTS method, which adeptly exploits interaction sequences to magnify its impact, maintaining significant disruptive capabilities even under stringent mixed-defense strategies.


The phenomena identified in this study not only underscore the elevated threat level posed by the DARTS framework, but also illuminate the urgent need for continuous innovation in defense strategies within Federated Sequential Recommendation (FSR) systems. These findings reveal how adversarial methods can exploit nuanced patterns and latent features within user-item interaction data to effectively circumvent conventional security measures.
Moreover, the conducted comparative analysis acts as a pivotal benchmark for understanding the magnitude of emerging challenges and the strategic evolution required for robust adversarial resilience. Such insights are instrumental in guiding the design of next-generation federated learning defenses that can adapt to increasingly sophisticated and covert attack paradigms.
\section{GenAI Usage Disclosure}
In this paper, we utilized generative AI to refine the text in the experimental section. Additionally, for code development, some portions of the code were assisted by generative AI, which also reviewed the code for bugs and modified certain functionalities.
\bibliographystyle{ACM-Reference-Format}

\clearpage

\end{document}